\documentclass[fleqn,usenatbib]{mnras}

\usepackage[T1]{fontenc}
\usepackage{ae,aecompl}
\usepackage{graphicx}	
\usepackage{amssymb}	
\usepackage{amsmath}
\usepackage{float}

\def\be{\begin{equation}}
\def\ee{\end{equation}}
\def\ba{\begin{eqnarray}}
\def\ea{\end{eqnarray}}

\title[]{Gravitational Wave Detection on OJ 287 via Pulsar Timing Array
}

\author[J. W. Chen and Y. Zhang]{
Jie-Wen Chen\thanks{E-mail: chjw@mail.ustc.edu.cn} and
Yang Zhang,\thanks{E-mail: yzh@ustc.edu.cn}
\\
Department of  Astronomy, Key Laboratory for Researches in Galaxies and Cosmology, \\
University of Science and Technology of China,   Hefei, Anhui, 230026,  China
}

\begin{document}
\label{firstpage}
\pagerange{\pageref{firstpage}--\pageref{lastpage}}

\maketitle

\begin{abstract}~

Blazar OJ 287 is a candidate of nano-Hertz gravitational wave (GW) source.
In this paper, we investigate the GW generated by OJ 287
and its potential detections through pulsar timing array (PTA).
Firstly, we obtain the orbit and the corresponding GW strain of OJ 287.
During the time span of next 10 years (2019-2029),
the GW  of OJ 287 will be active before 2021, with a peak strain amplitude $8\times 10^{-16}$,
and then decay after that.
When OJ 287 is silent in GW channel during 2021-2029,
the timing residual signals of the PTA will be dominated by the ``pulsar term" of the GW strain,
and it provides an opportunity to observed the ``pulsar term".
Furthermore, we choose 26 pulsars with white noises below 300 ns to detect the GW signal of OJ 287,
evaluating their timing residuals and signal-to-noise ratios (SNRs).
The total SNR (with cadence of 2 weeks in the next 10 years) of the PTA ranges from $1.9$ to $2.9$,
corresponding to a weak GW signal for current sensitivity level.
Subsequently, we investigate the potential measurements of the parameters of OJ 287 by the pulsars.
In particular, PSR J0437-4715 with precisely measured distance, has potential to constrain the polarization angle with uncertainty below $8^\circ$,
and the pulsar will play an important role in the future PTA observations.
\end{abstract}


\section{Introduction}~

Blazar OJ 287 is considered as a supermassive black hole (SMBH) binary system,
and its orbital evolution has been studied in detail in the works
\cite{Sillanpaa et al. 1988, Lehto and Valtonen 1996, Sundelius et al. 1997,
Valtonen and Ciprini 2012, Rampadarath Valtonen and Saunders 2007,
Valtonen  Ciprini and Lehto 2012,
Pihajoki  Valtonen and Ciprini 2013, Valtonen and Pihajoki 2014} and \cite{Valtonen et al. 2006, Valtonen et al. 2006b, Valtonen et al. 2008,
Valtonen et al. 2010a, Valtonen et al. 2010b, Valtonen et al. 2011,  Valtonen et al. 2014,  Valtonen et al. 2016}.
According to \cite{Valtonen Ciprini and Lehto 2012} and \cite{Valtonen et al. 2010a, Valtonen et al. 2016},
the mass of the primary is $1.84 \times 10^{10} M_{\odot}$,
mass of the secondary is  $1.4 \times 10^{8} M_{\odot}$,
the relative velocity between two black holes (BHs) is $\sim 0.1 c$,  with $c$ being the speed of light,
and the spin of the primary (dimensionless Kerr parameter) is $0.31$.

As a SMBH binary system, OJ 287 is also a candidate of gravitational wave (GW) source.
In our previous works \cite{Wu Zhang and Fu 2012, Zhang Wu  and Zhao 2013, Tong et al. 2013} and \cite{Cheng and Zhang 2013},
the GW strain of OJ 287 is evaluated in post-Newtonian (PN) approximation,
that it has frequency of $10^{-9}$ Hz and amplitude of $10^{-16}$.
The nano-Hertz frequency of the GW is not in the bands of the ground-based and space-based laser interferometers,
such as advanced LIGO \citep{Harry 2010, Aasi et al. 2015},
advanced Virgo \citep{Acernese et al. 2015},
Kagra \citep{Aso et al. 2013}
and eLISA \citep{Amaro-Seoane et al. 2012} \emph{etc.}.
However, the pulsar timing and pulsar timing array (PTA) observations,
with band from $10^{-9}$ Hz to $10^{-6}$ Hz,
can be used to detect this low-frequency GW signal \citep{Estabrook  and Wahlquist 1975, Detweiler 1979,
Foster and Backer 1990, Jenet et al. 2004, Hobbs et al. 2010, Wang Mohanty and Jenet 2014, Janssen et al. 2015,  Perera et al. 2018, Wang and Mohanty 2016}.
Hence, OJ 287 can serve as a target in the PTA projects,
such as the Parkes PTA (PPTA) \citep{Manchester et al. 2013, Hobbs 2013, Reardon et al. 2015},
the European PTA (EPTA) \citep{Kramer and Champion 2013, Desvignes et al. 2016}
and the North American Nanohertz Observatory for Gravitational Waves (NANOGrav) \citep{McLaughlin 2013, Arzoumanian et al. 2018},
or the International PTA (IPTA) as a collaboration of the aforementioned three regional PTAs \citep{Manchester 2013, Hobbs et al. 2010,
Verbiest et al. 2016},
and the next generation radio telescopes like SKA \citep{Smits et al. 2009, Janssen et al. 2015, Wang and Mohanty 2016}
and FAST \citep{Hobbs et al. 2014}.
The PTA data has been accumulated for over 10 years,
having found several pulsars with timing noise level below $1~\mu$s \citep{Reardon et al. 2015, Desvignes et al. 2016, Verbiest et al. 2016, Arzoumanian et al. 2015, Arzoumanian et al. 2018}.

The amplitude of GW strain
from general SMBH binary systems has been constrained to be smaller than $10^{-14}$
in \cite{Arzoumanian et al. 2014, Zhu et al. 2014, Babak et al. 2016} and \cite{Perera et al. 2018},
where OJ 287 is also under consideration in \cite{Arzoumanian et al. 2014, Zhu et al. 2014} and \cite{Babak et al. 2016}.
Furthermore, \cite{Gusev Porayko and Rudenko 2014} has given the upper limit of GW amplitude of OJ 287, which is also of the order $10^{-14}$.
In this paper, we shall use the PTA consisting of 26 pulsars with white noises below 300 ns,
mainly based on the recently released data in \cite{Arzoumanian et al. 2018},
to evaluate the timing residual signals and signal-to-noise ratios (SNRs) due to the GW.
If the PTA can detect the GW signature of OJ 287 in their timing residuals,
it will help to understand the structure of the SMBH binary system,
and improve the measurements of the parameters of OJ 287,
as a complementary to the optical observations.
Therefore, we shall also investigate the potential measurements of  parameters of OJ 287 by the PTA observations.

This paper is organized as follows.
In Section \ref{sec motion},
we shall reconstruct the orbital motion of OJ 287 based on PN approximation,
referring to the work in \cite{Valtonen et al. 2016}.
In Section \ref{GW},
we shall evaluate the GW strain of OJ 287 and the corresponding timing residual signals and SNRs of the PTA,
for the next 10 years.
In Section \ref{constrain},
we shall investigate the potential measurements of parameters of OJ 287 by the pulsars.
Finally, a summary will be given in Section \ref{conclusion}.

In this paper, we adopt the unit with gravitational constant $G=1$ and speed of light $c=1$.

\section{Orbit of OJ 287}\label{sec motion}
\subsection{Post-Newtonian Dynamics}\label{Post-Newtonian Dynamics}~

In this section, we shall reconstruct the orbit of OJ 287 in \cite{Valtonen et al. 2016},
in post-Newtonian (PN) approximation.
We take $(m_1, \vec x_1, \vec S_1)$ and $(m_2, \vec x_2, \vec S_2)$
as individual masses, spatial coordinates  in the center-of-mass frame and spins of two BHs respectively.
Note that in this paper subscript ``1" stands for the secondary and ``2" for the primary.
Since $m_1/m_2 \ll 1$ for OJ 287, the spin of the secondary $|\vec S_1| \sim m_1^2$
is negligible compared with spin of the primary $|\vec S_2| \sim m_2^2$ \citep{Valtonen et al. 2010a},
so we take $\vec S_1=0$ for simplicity.

The equation of motion for OJ 287 up to 3PN order, in the center-of-mass frame, is
 \begin{align}
 \label{equation of motion}
\frac{d^2\vec{x}}{dt^2} =\mathcal B_N+\mathcal B_{1PN}
+{\mathop \mathcal B\limits^{SO}}_{1.5PN}+\mathcal B_{2PN}
+{\mathop \mathcal B\limits^{Q}}_{2PN}    \nonumber \\
+\mathcal B_{2.5PN}
+{\mathop \mathcal B\limits^{SO}}_{2.5PN}+\mathcal B_{3PN}
+\mathcal O(c^{-7}),
\end{align}
where $\vec x=\vec x_1-\vec x_2$ is the relative coordinate, and
\[
\mathcal B_N=-\frac{m}{r^2}\vec e_r
\]
is the Newtonian gravitational acceleration,
with $m=m_1+m_2$ being the total mass of OJ 287,
$r=|\vec x|$ and $\vec e_r=\vec x/r$.
$\mathcal B_{1PN}$, $\mathcal B_{2PN}$, $\mathcal B_{2.5PN}$ and $\mathcal B_{3PN}$ are the
non-spin PN terms, where
\begin{align}
\mathcal B_{1PN}=-\frac{m}{r^2} \Big[
\big(-\frac{3\dot{r}^2\nu}{2}+v^2+3\nu v^2
-\frac{m}{r}(4+2\nu)\big)\vec e_r+ \nonumber \\
(-4\dot{r}+2\dot{r}\nu)\vec v \Big], \nonumber
\end{align}
with $\nu=m_1 m_2/m^2$,
$\vec v=d\vec x/dt$, $v=|\vec v|$,
and the formulations of other non-spin PN terms are given by \cite{Blanchet 2014}.
${\mathop \mathcal B\limits^{SO}}_{1.5PN}$ and ${\mathop \mathcal B\limits^{SO}}_{2.5PN}$
is the spin-orbit contributions, where
\begin{align}
{\mathop \mathcal B\limits^{SO}}_{1.5PN}=\frac{1}{r^3}
\{\vec e_r [12\vec S\cdot(\vec e_r \times \vec v)
+6\frac{\delta m}{m}\vec\Sigma\cdot(\vec n \times \vec v)] \nonumber \\
+9(\vec e_r \cdot \vec v)\vec e_r \times \vec S-
 7\vec v \times \vec S-3\frac{\delta m}{m} \vec v \times \vec \Sigma\}, \nonumber
\end{align}
with $\delta m=m_1-m_2$, $\vec S=\vec S_1+\vec S_2 \simeq \vec S_2$, $\vec \Sigma = \frac{m}{m_2}\vec S_2-\frac{m}{m_1}\vec S_1 \simeq \frac{m}{m_2}\vec S_2$ for OJ 287,
and the formulation of ${\mathop \mathcal B\limits^{SO}}_{2.5PN}$  can be found in
\cite{Kidder 1995} and \cite{Faye Blanchet and Buonanno 2010}.
One usually introduces the dimensionless Kerr parameter $\chi$ to
quantify the spin as $|\vec S_2|=\chi m_{2}^2$, and
$\chi$ takes value from $0$ for non-spin BH to $1$ for extreme Kerr BH.
Furthermore,
\begin{align}
\label{BQ}
{\mathop \mathcal B\limits^{Q}}_{2PN}
=-\frac{3 m Q}{2 m_2 r^4}
\{[5(\vec e_r \cdot \vec e_S)^2-1]\vec e_r-2(\vec e_r \cdot \vec e_S)\vec e_S\},
\end{align}
is the quadrupole-monopole term \citep{Barker and O'Connell 1975},
where $\vec e_S=\vec S/|\vec S|$ and $Q=-\vec S_2^2/m_2=- \chi^2 m_2^3$ is the quadrupole scalar of the primary \citep{Thorne 1980, Will 2008}.
It is seen that the equation of motion \eqref{equation of motion} does not
contain the spin-spin terms,
since we have neglected the spin of the secondary.

Synchronously, the spin of OJ 287 evolves following
\be
\label{spin precession}
\frac{d\vec S}{dt} =
  {\mathop \mathcal U \limits^{SO}}_{1PN}
+{\mathop \mathcal U \limits^{Q}}_{1.5PN}
+
{\mathop \mathcal U \limits^{SO}}_{2PN}
+\mathcal O(c^{-5}).
\ee
${\mathop \mathcal U \limits^{SO}}_{1PN}$ and ${\mathop \mathcal U \limits^{SO}}_{2PN}$ are the spin-orbit terms,
and their formulations 
are given in \cite{Faye Blanchet and Buonanno 2010}.
Furthermore, the quadrupole-monopole term
\begin{align}
\label{UQ}
{\mathop \mathcal U \limits^{Q}}_{1.5PN}=-\frac{3 m_1 Q}{r^3}
(\vec e_S\cdot \vec e_r)(\vec e_r \times \vec e_S) \nonumber \\
=\frac{3 m_1 |\vec S_2|^2}{m_2 r^3}
(\vec e_S \cdot \vec e_r)(\vec e_r\times \vec e_S)
\end{align}
is given by \cite{Barker O'Brien and  O'Connell 1981, D'Eath 1975} and \cite{Majar and Mikoczi 2012}.

In the following,
we shall reconstruct the orbit of OJ 287
from Eqs. \eqref{equation of motion} and \eqref{spin precession},
referring to the works in \cite{Lehto and Valtonen 1996, Sundelius et al. 1997, Valtonen  Ciprini and Lehto 2012} and \cite{ Valtonen et al. 2010a, Valtonen et al. 2016}.

\subsection{Orbit Reconstruction}\label{Orbit Reconstruction}~

To investigate the motion of OJ 287,
we shall at first establish a 3-dimensional coordinate system,
where the center-of-mass of OJ 287 is fixed at the origin.
Let the accretion disk of the primary be fixed on the (x,y)-plane,
and the direction of spin $\vec e_S$ point nearly along the z-direction \citep{Valtonen et al. 2010a, Valtonen et al. 2016}.
According to \cite{Lehto and Valtonen 1996, Sundelius et al. 1997} and \cite{Valtonen et al. 2010a},
the orbital plane of OJ 287 is taken to be nearly perpendicular to the accretion disk.
It is known that the orbital plane precesses when the spin is considered,
and we assume that the current (around the year 2019) orbit of the secondary is on the (x,z)-plane.

\begin{figure}
\includegraphics[width=0.4\textwidth]{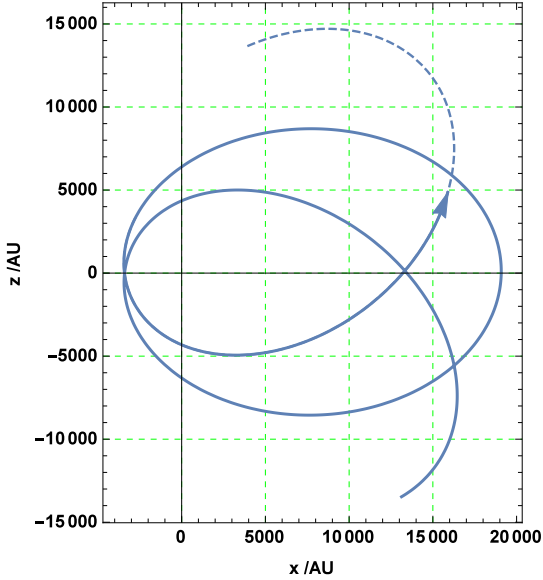}
\caption{
\label{orbit}
Orbit of the secondary of OJ 287 from the year 2000 to 2023 (solid curve), and 2023 to 2029 (dashed curve).}
\end{figure}

From Eqs. \eqref{equation of motion} and \eqref{spin precession}, we can obtain the orbit of OJ 287.
In our computation,
the masses of two BHs are taken to be $m_1=1.46 \times 10^8$ M$_{\odot}$ and $m_2=1.84 \times 10^{10}$ M$_{\odot}$ \citep{Valtonen Ciprini and Lehto 2012}, and spin of the primary is taken as $\chi=0.31$ \citep{Valtonen et al. 2016}.
Since $m_1\ll m_2$, the relative coordinate approximately represents the coordinate of the secondary, \emph{i.e.} $\vec x=\vec x_1$.
Then, by tuning the coordinates and velocities,
we reconstruct the orbit of the secondary of OJ 287 shown in Figure \ref{orbit}.
It is seen that the orbit from 2000 to 2023 in Figure \ref{orbit} is the same as that in \cite{Valtonen et al. 2016},
despite the difference in choosing coordinate system,
where the z-axis in this work is taken nearly along $\vec e_S$,
but along  $-\vec e_S$ in \cite{Valtonen et al. 2016}.
We also present some orbital parameters of OJ 287 in Table \ref{parameter},
where the orbital period and orbital eccentricity
are in agreement with the results in \cite{Valtonen et al. 2016}.
Furthermore, our computation implies that the secondary BH crosses the
accretion disk (z=0) in the years 2004.89, 2007.64, 2013.39, 2019.52 and 2021.36 respectively.
It is worth noting that the disk crossing leads to the optical outburst of OJ 287.
When the secondary crosses the disk, the disk gas is captured and shocked by the secondary,
and an outburst will happen after a time delay $t_d$ \citep{Lehto and Valtonen 1996}.
The time delay $t_d$ depends on the structure and  property of the accretion disk,
and can be written as a function of the separation between two BHs $R$ at the disk crossing moment, and other variables \citep{Lehto and Valtonen 1996}.
For simplicity, we only consider the dependance on $R$,
and take a simple empirical formulation,
concluded from \cite{Lehto and Valtonen 1996},
to estimate the time delay
\ba
\label{eqdelay}
\frac{t_{d}}{\rm 1 yr}= 4.753\times 10^{-13}\left(\frac{R}{\rm 1 AU} \right)^3
-3.294\times 10^{-9}\left(\frac{R}{\rm 1 AU} \right)^2 \nonumber \\
+2.068\times 10^{-5}\left(\frac{R}{\rm 1 AU} \right)-0.02428.
\ea
In our computation, the distances $R$ for all the disk crossings are $1.35\times 10^4$, $3.41\times 10^3$, $1.91\times 10^4$, $3.41\times 10^3$ and $1.32\times 10^4$ AU, and one from Eq. \eqref{eqdelay} can yield time delay $t_d=$ 0.81, 0.03, 2.47, 0.03 and 0.77 years respectively.
Hence, the outbursts should happen in the years 2005.7, 2007.6, 2015.9, 2019.6 (incoming) and 2022.1 (incoming),
consistent with the results in \cite{Valtonen et al. 2016}.
Due to the aspects above, the orbit of OJ287 in \cite{Valtonen et al. 2016} has been reconstructed with high accuracy.
\begin{table}
\caption{
\label{parameter}
Orbital parameters of OJ 287.
All time parameters in this table have multiplied the redshift factor $(1+z)$,
where $z=0.306$ for OJ 287 \citep{Lehto and Valtonen 1996}.
 }
\begin{center}
\begin{tabular}{|l|c|}
\hline
 orbital period & $12.0$ yr\\
 major axis &  $2.25 \times 10^4$ AU\\
 eccentricity &   $0.70$  \\
 period of major-axis precession & $1.14\times 10^2$ yr\\
 period of spin/orbital-plane precession & $2.68\times 10^3$ yr\\
\hline
\end{tabular}
\end{center}
\end{table}

\section{Gravitational Wave and Its Detection by PTA}\label{GW}
\subsection{GW Strain in PN Approximation}\label{Strain}~

With the orbit of OJ 287 obtained,
we shall evaluate the GW strain in the following.
The strain up to 2PN order is
\begin{align}
\label{waveform}
h^{ij}=\frac{2m\nu}{D}(Q^{ij}+P^{0.5}Q^{ij}+PQ^{ij}+P{\mathop Q \limits^{SO}}_{ij}+P^{1.5}Q^{ij} \nonumber \\
+P^{1.5}{\mathop Q \limits^{SO}}_{ij}+P^{2}Q^{ij}+P^{2}{\mathop Q \limits^{Q}}_{ij}),
\end{align}
where $``i, j"$ denote $``x, y, z"$, and $D=1.64\times 10^3$ Mpc is the luminosity distance of OJ 287
\citep{Wu Zhang and Fu 2012, Zhang Wu  and Zhao 2013, Tong et al. 2013, Cheng and Zhang 2013}.
The formulations of all terms on the right hand side of \eqref{waveform},
except quadrupole-monopole term $P^{2}{\mathop Q \limits^{Q}}_{ij}$,
are given by \cite{Kidder 1995} and \cite{Will and Wiseman 1996}.
The quadrupole-monopole term in the equation of orbital motion \eqref{equation of motion} is usually considered,
since it is important in examining the no-hair theorem
\citep{Carter 1970, Thorne and Hartle 1985, Valtonen et al. 2011, Will 2008, Johannsen and Psaltis 2010},
so we would like to include the corresponding term in the GW strain.
The formulation of $P^{2}{\mathop Q \limits^{Q}}_{ij}$ is not given by previous references,
and here we give
\newpage
\be
\label{PQ}
P^{2}{\mathop Q \limits^{Q}}_{ij}=-\frac{3m}{m_2^2 r^3}\{e^i e^j [\vec S^2-5(\vec e_r \cdot \vec S)^2]+2(\vec e_r \cdot \vec S)(e^i S^j+e^j S^i)\},
\ee
for the first time,
where $e^i$ and $S^i$ respectively denote the components of $\vec e_r$ and $\vec S$ on the $i$-direction.
A brief derivation of \eqref{PQ} is given in Appendix 1.

It is convenient to adopt the spherical coordinate system $(r, \theta, \phi)$ to analyze the GW strain,
since in the coordinate, the GW of OJ 287 propagates along the $\vec e_r$ direction.
Hence, one can take $(\vec e_\theta, \vec e_\phi)$ as a basis to describe the
polarization components \citep{Wagoner and Will 1976}
\begin{align}
h^{\theta \theta} = \frac{1}{2}(\cos^2\theta~ h^{xx}-\sin2\theta~ h^{xz}+\sin^2\theta~ h^{zz}-h^{yy}), \nonumber  \\
h^{\theta \phi} = \cos\theta~ h^{xy}-\sin\theta~ h^{yz}.
\end{align}
According to \cite{Valtonen et al. 2016}, the earth is nearly on the $\theta = 0$ or $180^\circ$ direction,
and in our subsequent evaluations, the earth is assumed to be on the $z$-direction ($\theta=0$).

\begin{figure}
\begin{center}
\includegraphics[width=0.45\textwidth]{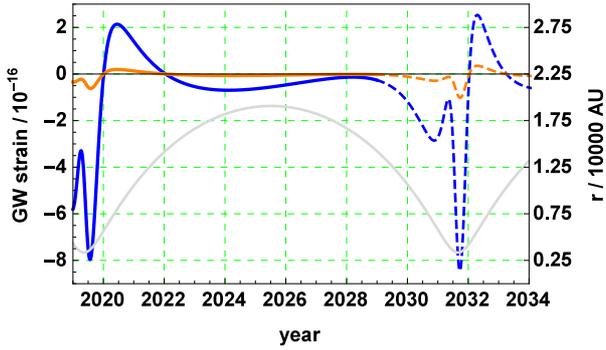}
\caption{The GW strain from OJ 287, in the local space around the earth, in the next 15 years.
The blue solid curve denotes the plus mode $h^{\theta \theta}$ from 2019 to 2029,
the blue dotted one $h^{\theta \theta}$ from 2029 to 2034,
the orange solid one denotes the cross mode $h^{\theta \phi}$ from 2019 to 2029,
and the orange dotted one $h^{\theta \phi}$ from 2029 to 2034.
The dark curve sketches the evolution of $r$, the distance between two BHs.
In this case, $\theta=0$ is taken. }
\label{h11}
\end{center}
\end{figure}

Given above, we can evaluate the GW strain of OJ 287, and
the result is sketch in Figure \ref{h11}.
It is seen that, the strain amplitude of the cross polarization mode $h^{\theta \phi}$
is much smaller than the  plus polarization mode $h^{\theta \theta}$.
This is because the orbit of OJ 287 is nearly ``edge-on" to the earth,
so the GW should almost have only one polarization mode \citep{Thorne 1980}.
Furthermore, the gravitational radiation will be active with peak value $|h^{\theta \theta}| \simeq 8\times 10^{-16}$ from 2019 to 2021 and from 2030 to 2033,
during which the secondary is around the orbital periastrons,
and OJ 287 will become silent in the GW channel in between 2022 to 2030.
The peak stain amplitude $8\times 10^{-16}$ is within the upper limit $10^{-14}$ in
\cite{Gusev Porayko and Rudenko 2014, Arzoumanian et al. 2014, Zhu et al. 2014} and \cite{Babak et al. 2016},
and consistent with the estimated value $\simeq 10^{-15}$ in \cite{Zhang Wu  and Zhao 2013, Tong et al. 2013} and \cite{Cheng and Zhang 2013}.
In addition, compared with the results in \cite{Zhang Wu  and Zhao 2013, Tong et al. 2013} and \cite{Cheng and Zhang 2013},
the result in Figure \ref{h11} is a developed version,  based on the accurately reconstructed orbit of OJ 287 in Figure \ref{orbit}.

\subsection{Timing Residuals of the PTA}\label{GW Detection by PTA}~

In the following, we shall apply PTA to detect the GW of OJ 287.
It is known that, pulsars make pulses with good periodicity,
and when GW propagates in the space,
it changes the physical distances between the earth and the pulsars,
resulting in changing the time-of-arrivals of the pulses.
The timing residual for a pulsar, due to GW is \citep{Estabrook  and Wahlquist 1975, Detweiler 1979, Jenet et al. 2004}
\be
\label{R(t)}
R(t)=\frac{1+\cos \mu}{2}[r_+(t)\cos(2\psi)+r_{\times}(t)\sin(2\psi)],
\ee
where $\mu$ is the opening angle between OJ 287 and the pulsar, with respect to the earth,
and $\psi$ is the polarization angle,
which is the angle between $\vec e_\theta$-direction
and the position vector of the pulsar, with respect to OJ 287, on the celestial sphere.
The parameters $\mu$ and $\psi$ are also illustrated in Appendix 2.
The function $r_{+,\times}(t)$ is
\be
\label{residualintegrate}
 r_{+,\times}(t)=\int^t _0 \left[h^e_{+,\times}(\tau)-h^p_{+,\times}(\tau-\Delta t) \, \right]d\tau,
\ee
where $h^e$ is the strain on the earth, namely the so-called ``earth term", and $h^p$ the ``pulsar term" \citep{Zhu et al. 2014, Zhu et al. 2016},
the polarization mode $h_{+,\times}$ denotes the aforementioned $h^{\theta \theta, \theta \phi}$,
and the geometric time delay in the pulsar term is
\be
\label{time delay}
\Delta t=(1-\cos \mu) \frac{d}{c},
\ee
with $d$ being the spatial separation between the earth and pulsar,  and $c$ being the speed of light.
\begin{table*}
\caption{  \label{pulsars}
Parameters of the 26 pulsars. The polarization angle $\psi-\psi_0$ and anglular separation $\mu$ are given by Eqs. \eqref{psi} and \eqref{mu} respectively in Appendix 2.
The white noise of the root-mean-squared timing residuals for the first 6 pulsars are cited from \citep{Zhu et al. 2016},
which broadly represents the level of the actual measurements \citep{Manchester et al. 2013, Arzoumanian et al. 2015, Desvignes et al. 2016, Verbiest et al. 2016}.
The white noises for the last 20 pulsars are from the recently measured data \citep{Arzoumanian et al. 2018}.
For some pulsars, the white noise is not released and should be smaller than the full noise,
so the ``white noise (ns)" column is written as ``<x", with ``x" being the full noise.
We note that ``J1939+2134" is also called ``B1937+21" \citep{Arzoumanian et al. 2018}. }
\begin{center}
\begin{tabular}{|l|c|c|c|c|l|c|c|c|c|}
\hline
name & $\psi-\psi_0$($^\circ$) & $\mu$ ($^\circ$) & $\Delta t$ (yr) & white noise (ns) &
name & $\psi-\psi_0$($^\circ$) & $\mu$ ($^\circ$) & $\Delta t$ (yr) & white noise (ns) \\
\hline
J0437-4715 & 52.2 & 88.7 & 499.5 & 58  & J1939+2134 &-114.8 & 134.4 & 19408.4 & 104 \\
J1640+2224 &-148.9 & 104.7 & 6219.5 & 158  & J1744-1134 & 176.9 & 133.593 & 2204.2 & 203 \\
J2317+1439 & 42.1 & 130.7 & 10990.6 & 267 & J2241-5236 & 64.0 & 141.5 & 5581.1 & 300 \\
\hline
J2234+0611 & 44.5 & 144.1 & 8972.0 & <30 & J1911+1347 & 128.8 & 137.7 & 7716.4 & <54 \\
J1909-3744 & 136.6 & 151.2 & 6976.8 & 70 & J2017+0603 & 110.3 & 152.3 & 8607.9 & <91 \\
J1741+1351 & 148.4 & 121.6 & 8896.3 & <102 & J1713+0747 & 158.4 & 118.9 & 5952.4 & 103 \\
J2043+1711 & 94.5 & 142.6 & 9129.7 & <119 & J0645+5158 & -59.6 & 40.5 & 625.4 & <180 \\
J1600-3053 & 154.2 & 113.9 & 8251.5 & <181 & J1614-2230 & 163.9 & 115.3 & 3259.4 & <183 \\
J1832-0836 & 166.7 & 143.8 & 4772.6 & <184 & J0740+6620 & -80.1 & 47.8 & 459.7 & <190 \\
J0613-0200 & 24.6 & 45.3 & 753.1 & 199 &  J2229+2643 & -62.9 & 127.8 & 9469.3 & <203 \\
J1853+1303 & 134.3 & 135.5 & 7373.9 & <205 & J2234+0944 & 48.2 & 141.3 & 9234.5 & <205 \\
J1923+2515 & -117.1 & 129.5 & 6402.4 & <229 & J0030+0451 & 19.3 & 121.5 & 1588.6 & 241 \\
J2010-1323 & 149.2 & 167.5 & 7477.2 & <260 & J1918-0642 & 153.4 & 153.1 & 6911.8 & <297 \\
\hline
\end{tabular}
\end{center}
\end{table*}

\begin{figure*}
\includegraphics[width=0.9\textwidth]{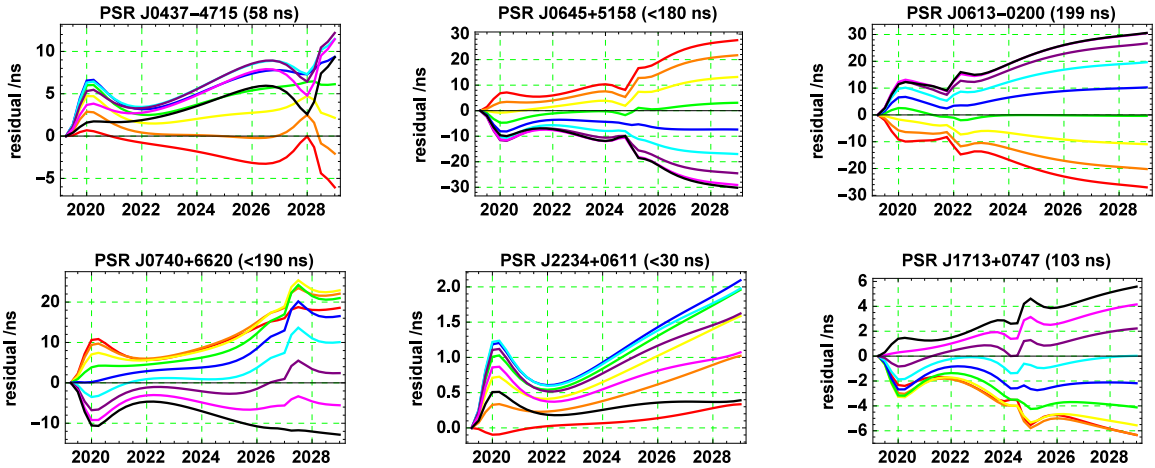}
\caption{The timing residuals due to the GW of OJ 287 of the ``best" 6 pulsars, in the next 10 years.
The colored curves correspond to various values of $\psi_0$, in detail,
$\psi_0=0$ (red), $\psi_0=10^{\circ}$ (orange), $\psi_0=20^{\circ}$ (yellow), $\psi_0=30^{\circ}$ (green),
$\psi_0=40^{\circ}$ (blue), $\psi_0=50^{\circ}$ (cyan), $\psi_0=60^{\circ}$ (purple), $\psi_0=70^{\circ}$ (magenta) and $\psi_0=80^{\circ}$ (black).
The residuals for other pulsars are given in Figure \ref{residual2} in Appendix 3.
}
\label{residual}
\end{figure*}

Since the GW of OJ 287 has not been detected yet, we can treat it as a weak signal.
For weak signal, one can assume that the white noise of pulsar is much larger than the red noise,
dominating the total noise \citep{Siemens et al. 2013, Janssen et al. 2015}.
Hence in the following evaluations, we only consider the white noises of the pulsars.
To detect the GW of OJ 287,
we select 26 pulsars,
with white noises below $300$ ns.
The cutoff $300$ ns here is arbitrarily chosen as in \cite{Zhu et al. 2016}.
The locations for pulsars are given in Table \ref{position} in Appendix 2,
and based on the position, we have calculated the parameters $\psi$, $\mu$ and $\Delta t$  from Eqs. \eqref{mu}, \eqref{psi} and \eqref{time delay} respectively,
and shown the results in Table \ref{pulsars}.
It is seen that there is an undetermined constant $\psi_0$  in $\psi$,
and this is because the observer on the earth may choose a basis different from $(\vec e_\theta, \vec e_\phi)$
to describe the polarization modes, and here $\psi_0$ is the angle between $\vec e_\theta$ and the latitude-direction on the celestial sphere (see Appendix 2).
In this paper, we take the polarization angular constant $\psi_0$ as a parameter, free from the pulsars.
From Eq. \eqref{R(t)}, it is seen that $R(t) \rightarrow -R(t)$ after the transformation $\psi\rightarrow \psi+90^\circ$,
so we only need to consider the value of $\psi_0$ ranging from $0$ to $90^\circ$.
Table  \ref{pulsars} also gives the white noises of the root-mean-squared timing residuals of the pulsars.
We mention that the white noises for 6 pulsars are quoted from \citep{Zhu et al. 2016},
which is based on simulated data sets, but represents the level of actual measurements in \cite{Manchester et al. 2013, Arzoumanian et al. 2015, Desvignes et al. 2016} and \cite{Verbiest et al. 2016}.
Furthermore, for some pulsars, the white noises are not released and should be smaller than the full noises,
so we shall use their full noises to conservatively estimate the levels of the white noises in Table  \ref{pulsars}.

\begin{figure*}
\begin{center}
\includegraphics[width=0.9\textwidth]{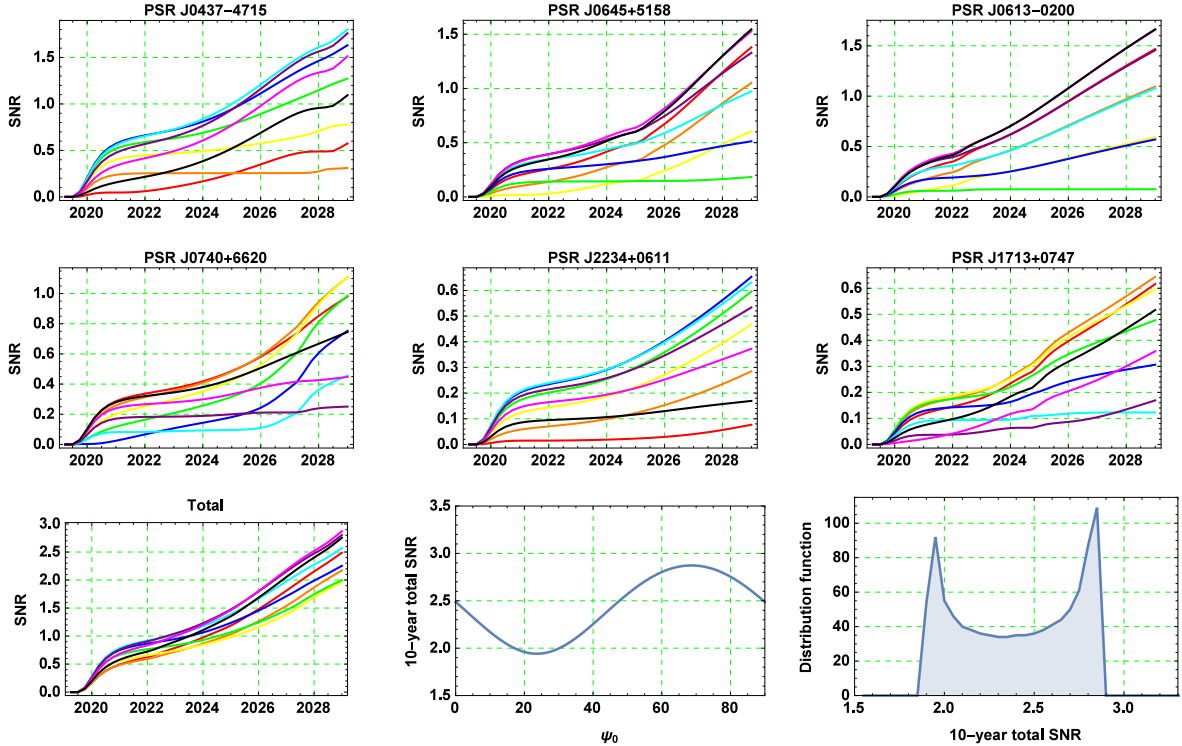}
\caption{The accumulations of individual SNRs for the best 6 pulsars and the total SNR of the 26 pulsars, in the next 10 years.
The colored curves correspond to various $\psi_0$, as is demonstrated in Figure \ref{residual}.
Panel (row 3, column 2) sketches the 10-year total SNR as a function of $\psi_0$,
Panel (row 3, column 3) presents the distribution function of total SNR, assuming the value of $\psi_0$ is evenly distributed.
}
\label{SNR fig}
\end{center}
\end{figure*}

By taking the values of the parameters $\mu$, $\psi$ and $\Delta t$ in Table \ref{pulsars} into Eq. \eqref{residualintegrate},
we can evaluate the timing residual signals of the pulsars due to the GW of OJ 287.
In the calculation, $\psi_0$ is taken as a parameter ranging from 0 to $90^\circ$.
At first, one needs to obtain the orbit of OJ 287 from Eq. \eqref{equation of motion}
and the GW strain from Eq. \eqref{waveform}, for a very long time span,
containing the geometric time delays $\Delta t$ of all pulsars.
In the following analysis, we consider the residuals in the time span of the next 10 years (2019-2029).
During the time span, OJ 287 will be active before 2021 and silent after that, in the GW channel.
The resulting residuals for the best 6 pulsars are sketched in Figure \ref{residual},
and for the rest ones are given in Figure \ref{residual2} in Appendix 3.
Here the ``best" is evaluated by that the pulsars have maximum values of timing residual-to-white noise ratios.
Considering both Figures \ref{residual} and \ref{residual2},
it is seen that the timing residuals generally range from $10^{-10}$ to $10^{-8}$  seconds,
and are smaller than the white noise levels by 2 orders of magnitude for most pulsars.
But for the pulsars in Figure \ref{residual}, the timing residual-to-white noise ratios can be close to or even larger than 0.1,
for certain values of $\psi_0$.
Hence by long-time observations,
it is possible to increase the SNRs for these pulsars to be larger than 1,
which will be discussed in Section \ref{Signal-to-Noise Ratios of the PTA}.
Furthermore, according to Eq. \eqref{residualintegrate},
if the GW is weak in the space between the earth and the pulsar,
the residual signal should stop evolving.
However, it is seen from Figures \ref{residual} and \ref{residual2} that the evolutions of the timing residuals will not freeze after 2021,
when OJ 287 is silent in GW channel.
This is because the timing residual is not only related to the earth term $h^e$,
but also to the pulsar term $h^p$,
and in this case $|h^e(\tau)|$ is small but $|h^p(\tau-\Delta t)|$ is large.
Therefore, the timing residuals will be dominated by $h^p$ after 2021,
and it is a good chance to observe the pulsar term in that duration.
We note that the distances of most pulsars in Table \ref{position} are poorly measured,
and the observation of pulsar term can improve the pulsar distance $d$ estimation,
to the level that the uncertainty of the geometric time delay $\Delta t=(1-\cos \mu) d/c$
is smaller than the orbital period of OJ 287 \citep{Lee et al. 2011, Zhu et al. 2016}.
The measurement of the pulsar term can also help to probe the binary evolutionary histories \citep{Mingarelli et al. 2012}.

\subsection{Signal-to-Noise Ratios of the PTA}\label{Signal-to-Noise Ratios of the PTA}~

To better evaluate the probability of the detection,
we calculate the SNRs of the GW signal in the timing residuals of the pulsars, in the next 10 years,
which is defined as \citep{Wang Mohanty and Jenet 2014, Zhu et al. 2016, Wang and Mohanty 2016}
\be
\label{SNR}
\rho^2={\mathop \sum \limits_{i}^{26}} \rho_i^2={\mathop \sum \limits_{i}^{26}}{\mathop \sum \limits_{j}^{N_i}} \frac{R^2(t_j)}{\sigma_i^2(t_j)},
\ee
where $\rho$ is the total SNR of the PTA,
and $\rho_i$, $N_i$ and $\sigma_i$ denote the individual SNR, number of data points and noise of the i-th pulsar, respectively.
The value of $\sigma_i$ can be estimated by the white noise in Table \ref{pulsars},
and hence the $R(t_j)/\sigma_i(t_j)$ term on the right hand side of Eq. \eqref{SNR} is the aforementioned timing residual-to-white noise ratio.
Furthermore, we assume that the observation is equal-time-interval,
and take the typical cadence of $\Delta T=2$ weeks \citep{Verbiest et al. 2016},
so  the total number of data points for each pulsar, in the time span of $T=10$ years, is $N_i=T/\Delta T=261$.
As $N_i \gg 1$, the summation of data points in Eq. \eqref{SNR} can be approximately evaluated by an integrate
\be
\rho^2_i= \int \frac{dt}{\Delta T} \frac{R^2(t)}{\sigma_i^2}.
\ee

The individual SNRs for the best 6 pulsars and total SNR of the PTA are presented in Figure \ref{SNR fig},
and the SNRs for other pulsars are shown in Figure \ref{SNR fig2} in Appendix 3.
The ``best" here denotes the largest SNRs,
and the best 6 pulsars in Figure \ref{SNR fig} are consistent with those  Figure \ref{residual}.
One can see that the total SNR $\rho$ ranges from 1.9 to 2.9, depending on the value of $\psi_0$,
and its distribution function has a double-peak profile.
 The signal with $\rho \leq 2.9$ is generally deemed as weak signal in actual measurements \citep{Wang Mohanty and Jenet 2014, Zhu et al. 2016, Wang and Mohanty 2016},
so the GW of OJ 287 cannot be detected with high confidence at current PTA sensitivity level.
However, for pulsars like J0645+5158, J0740+6620 and J2234+0611 etc.,
the white noises are taken as the full noises,
thus their real SNRs are conservatively estimated,
and the real total SNR should be larger than the evaluated result ranging from 1.9 to 2.9.

\begin{figure*}
\begin{center}
\includegraphics[width=0.9\textwidth]{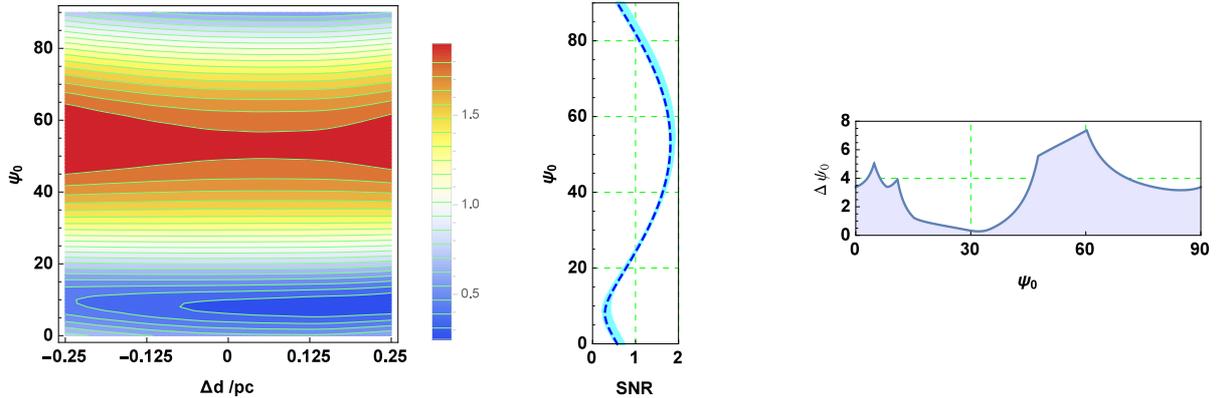}
\caption{Left: The contour of SNR for PSR J0437-4715, distributed from the distance uncertainty of $\Delta d$ and polarization angle constant $\psi_0$.
Middle: The dependence of $\psi_0$ on the measrued SNR. The cyan shadowed region takes into account all $|\Delta d|<0.25$ pc,
and the blue dashed curve corresponds to the case $\Delta d=0$.
Right: The measurement uncertainty of the polarization angle constant $\Delta \psi_0$, as a function of the estimated value of $\psi_0$.
}
\label{d-psi}
\end{center}
\end{figure*}

The next-generation telescope such as SKA can improve the PTA sensitivities by 2 orders of magnitude from the current IPTA level \citep{Janssen et al. 2015},
which implies that GW signal of OJ 287 can be detected with high total SNR $\rho \sim 10^2$ in the future SKA era \citep{Wang and Mohanty 2016}.

\section{Parameter Constraints on OJ 287}\label{constrain}~

If the GW of OJ 287 can be detected by the PTA in the next 10 years,
it can help to understand the structures and physical properties of OJ 287,
as a complementarity to the optical observations.
In this section, we shall apply these pulsars to measure,
or at least constrain, the parameters of OJ 287,
assuming the GW signal having been detected.

The measurement of the polarization angle constant $\psi_0$ is an important issue to be solved,
and we are also interested in the measurements of spin and quadrupole-monopole scalar of OJ 287.
The measurements  of these parameters are discussed in the following.

\subsection{Polarization Angle Constant $\psi_0$}~

As is demonstrated in the above,
the timing residuals and SNRs of the pulsars vary from polarization angle constant $\psi_0$,
which is totally undetermined.
We also have mentioned that the distance $d$ is poorly measured for most pulsars,
with uncertainty $|\Delta d|>0.1$ kpc \citep{Manchester et al. 2005},
corresponding to the uncertainty of geometric time delay $\Delta t$ much larger than 12 years, the orbital period of OJ 287.
The variation of the value of SNR, namely $\Delta \rho$, due to $\Delta d$,
should be generally comparable and degenerate to $\Delta \rho$ due to $\psi_0$,
making the measurements of $\psi_0$ difficult.
Therefore, to effectively measure or constrain the value of $\psi_0$,
pulsars with precisely measured distances are required.

PSR J0437-4715 is such a candidate, with the measured distance $d=0.15679\pm0.00025$ kpc \citep{Reardon et al. 2015}.
The small distance uncertainty $|\Delta d|=0.25$ pc corresponds to the uncertainty $0.8$ years of geometric time delay $\Delta t=499.5$ years,
and is much smaller than the orbital period $12$ years of OJ 287.
If the value of SNR depends on $\psi_0$ more strongly than on $\Delta d$, $\psi_0$ can be constrained precisely in this case.
Hence, we calculate the SNR distribution in the ($\Delta d$, $\psi_0$)-space for PSR J0437-4715 in left panel of Figure \ref{d-psi}.
It is clear that the value of SNR generally varies more strongly along the $\psi_0$-direction,
so $\psi_0$ should be able to be stringently constrained by PSR J0437-4715.
Furthermore, we present an overall constraint of $\psi_0$ as a function of the observed SNR, considering all $|\Delta d|<0.25$ pc,
in the middle panel of Figure \ref{d-psi}.
An obvious correspondence relation between $\psi_0$ and SNR is seen, and the dispersion of the relation due to $\Delta d$ is small.
The width of the cyan shadowed region along the vertical direction in the middle panel represents the measurement uncertainty of the
polarization angle constant $\Delta \psi_0$.
The measurement uncertainty $\Delta \psi_0$ as a function of the estimated value of $\psi_0$ is illustrated in the right panel of Figure \ref{d-psi}.
It is seen that the value of $\Delta \psi_0$ is generally smaller than $8^\circ$,
and the most precise measurement occurs around $\psi_0=30^\circ$ with uncertainty $\Delta \psi_0<1^\circ$.
Therefore, the polarization angle constant $\psi_0$, can be measured or constrained with high precision by PSR J0437-4715,
when the GW of OJ 287 is detected in the future.

After measuring or constraining the value of $\psi_0$ by PSR J0437-4715,
one can improve pulsar term observations and the distance measurements for other pulsars, as is mentioned in Section \ref{GW Detection by PTA}.
Hence PSR J0437-4715 plays a key role in the detection.

\subsection{Spin and Quadrupole Scalar}\label{Spin and Quadrupole Momentum}~

In this subsection, we shall evaluate the precisions in the potential measurements of
the spin and quadrupole scalar of OJ 287,  in the PTA observations.

As is shown in Table \ref{parameter}, the spin of OJ 287 is a long-term effect,
so it should be not significant in the orbit and GW strain during the next 10 years.
However, due to the long geometric time delays $\Delta t$ of the pulsars,
which are generally comparable to or larger than the period of spin precession $2.68\times 10^3$ years,
the spin should affect the timing residuals of pulsars strongly,
and can be constrained precisely.
For simplicity, we fix the spin direction and only investigate the absolute value of the spin $|\vec S_2|$,
or equivalently, the dimensionless Kerr parameter $\chi$.
In \cite{Valtonen et al. 2016}, by fitting the observed outbursts of OJ 287 as discussed in Section \ref{Orbit Reconstruction},
the orbit is precisely constrained and the spin is measured as $\chi=0.31\pm0.01$.
The result represents the current level of optical measurement.

The quadrupole-monopole effect is a minor term in the spin effect.
Although the effect is weak, it is important in examining the no-hair
theorem \citep{Carter 1970, Thorne and Hartle 1985, Valtonen et al. 2011, Will 2008, Johannsen and Psaltis 2010},
as is mentioned in Section \ref{Strain}.
We have introduced the quadrupole scalar $Q=- \chi^2 m_2^3$ in Section \ref{Post-Newtonian Dynamics},
and now let us redefine it as $Q=- q \chi^2 m_2^3$,
where $q$ is a dimensionless parameter, with
$q=1$ corresponding to the case that the no-hair theorem holds.
Synchronously, the quadrupole-monopole strain $P^{2}{\mathop Q \limits^{Q}}_{ij}$ in \eqref{PQ} should also be modified as in Eq. \eqref{PQ3} (see Appendix 1).
Similarly to the measurement of spin,  $q=1.0\pm0.3$ is obtained in \cite{Valtonen et al. 2011}.

If a parameter $f$ is measured as $f=f_0\pm \Delta f$ in the PTA observations,
the measurement uncertainty $\Delta f$ should contribute to
the uncertainty of the individual SNR of the $i$-th pulsar,
$\Delta \rho_i=|\rho_i(f_0+\Delta f)-\rho_i(f_0)|$.
$\Delta \rho_i$ represents the resolution of the measured SNR value $\rho_i$,
hence one can take
\be
\label{Delta SNR}
\Delta \rho_i=\epsilon_i,
\ee
where $\epsilon_i$ is a small constant.
By finding the minimal solution of Eq. \eqref{Delta SNR},
one can obtain the measurement uncertainty $\Delta f$.

In the following, we shall evaluate the measurement uncertainties of the spin $\Delta \chi$, and of quadrupole scalar $\Delta q$ in the PTA observations.
Firstly, we assume that the GW of OJ 287 will be detected in the next 10 years,
although it seems difficult for the current PTA sensitivities.
In addition, the pulsar distances are assumed to be precisely measured,
the polarization angle constant is fixed as $\psi_0=0$,
and the estimated values of spin and quadrupole scalar are $\chi_0=0.31$ and $q_0=1$ respectively,
as the results in the optical measurements \citep{Valtonen et al. 2011, Valtonen et al. 2016}.
In the case of $\psi_0=0$, the best six pulsars, with largest individual SNRs, are
J0613-0200, J0645+5158, J0740+6620, J0437-4715, J1640+2224 and J1614-2230 respectively
 (not all the ones in Figs. \ref{residual} and \ref{SNR fig}).
For simplicity, we only calculate the  measurements uncertainties from the six pulsars above.
By solving Eq. \eqref{Delta SNR}, the resulting $\Delta \chi$ and $\Delta q$ are given in Table \ref{uncertainty}.

It is seen that the measurement uncertainties from the pulsars are comparable or even smaller than those in the optical measurements,
with the resolution being arbitrarily taken as $\epsilon_i=0.01$.
This implies that the GW observations by PTA can be complementary to the optical observations.
If the individual SNR can be increased,
e.g. in the future SKA era,
the resulting uncertainties should be smaller, corresponding to higher precisions.
Furthermore, it is clear that $\Delta \chi$ and $\Delta q$ have dependence on the geometric time delay $\Delta t$.
For pulsars with $\Delta t>2.56\times 10^3$ years, the period of spin precession of OJ 287,
the values of $\Delta \chi$ and $\Delta q$ are much smaller than those of
the pulsars with $\Delta t<2.56\times 10^3$ years.
 As is mentioned above, the measurement on quadrupole scalar $q$ can examine the no-hair theorem,
and the results in Table \ref{uncertainty} imply that the PTA observation has potential to examine the theorem at the high precision of $10^{-16}$.

\begin{table}
\caption{  \label{uncertainty}
Measurement uncertainties of spin $\Delta \chi$ at the estimated value $\chi_0=0.31$,
and of quadrupole scalar $\Delta q$ at $q_0=1.0$ for the best six pulsars.
In the simulation, $\psi_0=0$ is taken and the resolution of the measured SNR is taken as $\epsilon_i=0.01$. }
\begin{center}
\begin{tabular}{|l|c|c|}
\hline
name ($\Delta t $ / years) & $|\Delta \chi|$ & $|\Delta q|$ \\
\hline
optical measurement   &    0.01    &  0.3  \\
\hline
J0740+6620 (459.7) & <0.003 & <0.7  \\
J0437-4715 (499.5) &  0.004 & 0.4  \\
J0645+5158 (459.7) &  <0.002 & <0.3  \\
J0613-0200 (753.1) & 0.004  & 0.4  \\
J1614-2230 (3259.4) & <$1\times 10^{-16}$ & <$3\times 10^{-16}$ \\
J1713+0747 (5952.4) &  $3\times 10^{-17}$  & $1\times 10^{-16}$  \\
\hline
\end{tabular}
\end{center}
\end{table}

\section{Conclusion and Discussion}\label{conclusion}~

In this paper, we have investigated the GW of OJ 287 and its potential detection by the PTA.

At first, we have reconstructed the orbit in \cite{Valtonen et al. 2016} with high accuracy.
The orbit is evaluated in Post-Newtonian approximation up to the 3PN order,
and the simplified approximations such as non-evolving orbit, circular orbit and even the eccentric orbit \emph{etc.},
which are usually taken in the PTA analysis \citep{Zhu et al. 2014, Zhu et al. 2016, Janssen et al. 2015, Wang and Mohanty 2016, Perera et al. 2018}, are not used in this work.
Therefore, the calculated GW strain should be very close to the real one.

Then we have provided a forecast of the GW of OJ 287 in the next 10 years (2019-2029).
The GW will be active in the years 2019-2021, with a peak strain amplitude $8\times 10^{-16}$, and then decay after 2021.
The peak amplitude $8\times 10^{-16}$ is below the current PTA sensitivities \citep{Arzoumanian et al. 2014, Zhu et al. 2014, Gusev Porayko and Rudenko 2014, Babak et al. 2016, Perera et al. 2018},
but stronger than the levels of the next generation  PTA projects, like SKA \citep{Janssen et al. 2015, Wang and Mohanty 2016}.

The observation of the pulsar term $h^p$ of the strain is important in tracing the evolutionary histories of OJ 287 and improving the measurements of pulsar distances \citep{Mingarelli et al. 2012, Lee et al. 2011, Zhu et al. 2016}.
OJ 287 just provides such an opportunity, that the earth term will be weak in 2021-2029 as is forecasted above,
and then the timing residual signals of the PTA will be dominated by the ``pulsar term".

The quadrupole-monopole term is introduced in the GW strain for the first time.
Since the value of BH quadrupole scalar is closely related to the no-hair theorem \citep{Carter 1970, Thorne and Hartle 1985, Will 2008},
the observation of the quadrupole-monopole strain by PTA provides a method in examining the theorem in the GW window.
As is discussed in Section \ref{Spin and Quadrupole Momentum},
it is possible that the PTA observation can examining the theorem, at the uncertainty level of $10^{-16}$,  in the future SKA era.

Furthermore, we have evaluated the timing residuals and SNRs of 26 pulsars in the next 10 years.
The total SNR of the PTA ranges from 1.9 to 2.9.
The result $\rho \leq 2.9$ corresponds to  weak signal in actual detection \citep{Wang Mohanty and Jenet 2014, Zhu et al. 2016, Wang and Mohanty 2016},
but $\rho >1$ is already a happy news for theorists, since it seems not far away from detecting the signal.
Note that the resulting SNR above is conservatively estimated, and its real value should be larger than 1.9 to 2.9.
Additionally, in the future SKA era, the total SNR should be increased to $\rho \sim 10^2$ \citep{Janssen et al. 2015, Wang and Mohanty 2016}.
Given above, although the GW signal of OJ287 is weak for current sensitivity level,
it still has potential to be detected in the future.

In our evaluations, PSR J0437-4715 provides a great contribution in the total SNR, as is shown in Figure \ref{SNR fig}.
Moreover, its precisely measured distance $d=0.15679\pm0.00025$ kpc can potentially give a stringent constraint on
the polarization angle constant $\psi_0$, with precision $\Delta \psi_0<8^\circ$.
If the value of $\psi_0$ is fixed, it will furthermore improve the  pulsar term observations and the distance measurements for other pulsars,
when the GW signal is detected by these pulsars.
Due to the aspects above,
PSR J0437-4715 will play an important role in the future PTA detection.

At last, we have investigated the potential measurements of spin and quadrupole scalar of OJ 287 by PTA observations.
assuming the GW signal can be detected.
The resulting measurement uncertainties in Table \ref{uncertainty} are comparable or even smaller than the optical results in \cite{Valtonen et al. 2011} and
\cite{Valtonen et al. 2016}.
In particular, for the pulsars with long geometric time delays $\Delta t$,
larger than the period of spin precession of OJ 287,
the measurement precisions are very high.
In the future SKA era, when the individual SNRs for these pulsars are improved,
the uncertainties will be smaller.

\section*{Acknowledgements}
Y. Zhang is supported by NSFC Grant No.11275187, NSFC 11421303,
SRFDP, and CAS,
the Strategic Priority Research Program
``The Emergence of Cosmological Structures"
of the Chinese Academy of Sciences, Grant No. XDB09000000.

\section*{Appendix 1: Quadrupole-monopole Strain}~

The GW strain is determined by the equations of motion \eqref{equation of motion} and spin precession \eqref{spin precession}.
Previous references such as \cite{Kidder 1995} and \cite{Will and Wiseman 1996} do not give a $P^{2}{\mathop Q \limits^{Q}}_{ij}$ term in the GW strain
because they do not include the quadrupole-monopole terms in the equations of motion and spin precession.
In this work, we should give a $P^{2}{\mathop Q \limits^{Q}}_{ij}$ term in the GW strain.

We notice that there is a similarity between the quadrupole-monopole terms and the spin-spin terms in both the equation of motion and of spin precession.
For example, the 2PN-order spin-spin term in equation of motion is
\begin{align}
\label{SS}
{\mathop \mathcal B \limits^{SS}}_{2PN}= -\frac{3m}{m_1 m_2 r^4}
[\vec e_r(\vec S_1\cdot \vec S_2)
+\vec S_1(\vec e_r \cdot \vec S_2)         \nonumber \\
+\vec S_2(\vec e_r \cdot \vec S_1)-5\vec e_r(\vec e_r \cdot \vec S_1)(\vec e_r \cdot \vec S_2)],
\end{align}
given by \cite{Kidder 1995}.
By comparing above equation and \eqref{BQ},
we find that
\be
\label{QvsSS}
{\mathop \mathcal B\limits^{Q}}_{2PN}
    =\frac{m_1}{2m_2}{\mathop \mathcal B \limits^{SS}}_{2PN}|_{\vec S_1=\vec S_2=\vec S}.
\ee
We explain the relation \eqref{QvsSS} that the quadrupole momentum is a high-order effect of spin,
and it can be considered as a special self-coupling of spin-spin,
\emph{i.e.} a coupling between $\vec S_2$ and $\vec S_2$,
hence one has $\vec S_1=\vec S_2$.
The factor $1/2$ in \eqref{QvsSS} identifies the ``self-coupling".
Then the quadrupole momentum couples with the monopole (mass) $m_1$,
which is the origin of the coefficient $m_1$ in \eqref{QvsSS}.
Similarly, comparing the 1.5PN-order spin-spin term in equation of spin precession
\begin{align}
{\mathop \mathcal U \limits^{SS}}_{1.5PN}
=\frac{3|\vec S_1||\vec S_2|}{r^3}[(\vec e_r \cdot \hat S_1)(\vec e_r \times \hat S_2) \nonumber \\
+(\vec e_r \cdot \hat S_2)(\vec e_r \times \hat S_1)],
\end{align}
also given by \cite{Kidder 1995},
and Eq. \eqref{UQ},
one yields
\be
\label{QvsSS2}
{\mathop \mathcal U \limits^{Q}}_{1.5PN}
=\frac{m_1}{2m_2}{\mathop \mathcal U \limits^{SS}}_{1.5PN}|_{\vec S_1=\vec S_2=\vec S}.
\ee

Since the quadrupole-monopole term can be considered as a spin-spin term of self-coupling,
we can infer that the  relation should also appear in the strain as
\be
\label{QvsSS3}
P^{2}{\mathop Q \limits^{Q}}_{ij}=\frac{m_1}{2m_2}P^{2}{\mathop Q \limits^{SS}}_{ij}|_{\vec S_1=\vec S_2=\vec S}.
\ee
As $P^{2}{\mathop Q \limits^{SS}}_{ij}$ is given by \citep{Kidder 1995, Will and Wiseman 1996},
and one finally obtains
\[
P^{2}{\mathop Q \limits^{Q}}_{ij}=-\frac{3m}{m_2^2 r^3}\{e^i e^j [\vec S^2-5(\vec e_r \cdot \vec S)^2]+2(\vec e_r \cdot \vec S)(e^i S^j+e^j S^i)\},
\]
which is Eq. \eqref{PQ} in Section \ref{Strain}.

Note that, the results above are under the condition that the no-hair theorem holds.
If we release the condition as in Section \ref{Spin and Quadrupole Momentum},
taking the redefined quadrupole scalar $Q=-q \chi m_2^3$ into Eqs. \eqref{BQ} and \eqref{UQ},
the coefficient $m_1/2m_2$ should be replaced by $q m_1/2m_2$ in Eqs.  \eqref{QvsSS} and \eqref{QvsSS2},
and in \eqref{QvsSS3} as well.
Therefore, the GW strain in \eqref{PQ} should be modified to
\be
\label{PQ3}
P^{2}{\mathop Q \limits^{Q}}_{ij}=-\frac{3 q m}{m_2^2 r^3}\{e^i e^j [\vec S^2-5(\vec e_r \cdot \vec S)^2]+2(\vec e_r \cdot \vec S)(e^i S^j+e^j S^i)\}.
\ee

\begin{table*}
\caption{  \label{position}
Spatial locations of OJ 287 and the pulsars, which can be  found in the ATNF Pulsar Catalogue (http://www.atnf.csiro.au/research/pulsar/psrcat/) \citep{Manchester et al. 2005}. The precisely measured distance of PSR J0437-4715 is given by \citep{Reardon et al. 2015}.}
\begin{tabular}{|l|c|c|c|l|c|c|c|}
\hline
name & Right Ascension & Declination & d (kpc) & name & Right Ascension & Declination & d (kpc) \\
\hline
OJ 287 & 08:54:49 & +20:06:31 & $z=0.306$  \\
\hline
J0437-4715 & 04:37:16 & -47:15:09 & $0.15679\pm0.00025$ & J1939+2134 & 19:39:39 & +21:34:59 & 3.50 \\
J1640+2224 & 16:40:17 & +22:24:09 & 1.52 & J1744-1134 & 17:44:29 & -11:34:55 & 0.40 \\
J2317+1439 & 23:17:09 & +14:39:31 & 2.04 & J2241-5236 & 22:41:42 & -52:36:36 & 0.96 \\
J2234+0611 & 22:34:23 & +06:11:29 & 1.52 & J1911+1347 & 19:11:55 & +13:47:34 & 1.36  \\
J1909-3744 & 19:09:47 & -37:44:14 & 1.14 & J2017+0603 & 20:17:23 & +06:03:05 & 1.40 \\
J1741+1351 & 17:41:31 & +13:51:44 & 1.79 & J1713+0747 & 17:13:50 & +07:47:37 & 1.23 \\
J2043+1711 & 20:43:20 & +17:11:29 & 1.56 & J0645+5158 & 06:45:59 & +51:58:15 & 0.80 \\
J1600-3053 & 16:00:51 & -30:53:49 & 1.80 & J1614-2230 & 16:14:37 & -22:30:31 & 0.70 \\
J1832-0836 & 18:32:28 & -08:36:55 & 0.81 & J0740+6620 & 07:40:46 & +66:20:34 & 0.43 \\
J0613-0200 & 06:13:44 & -02:00:47 & 0.78 & J2229+2643 & 22:29:51 & +26:43:58 & 1.80\\
J1853+1303 & 18:53:57 & +13:03:44 & 1.32 & J2234+0944 & 22:34:46 & +09:44:30 & 1.59 \\
J1923+2515 & 19:23:22 & +25:15:41 & 1.20 & J0030+0451 & 00:30:27 & +04:51:40 & 0.32 \\
J2010-1323 & 20:10:46 & -13:23:56 & 1.16 & J1918-0642 & 19:18:48 & -06:42:35 & 1.12 \\
\hline
\end{tabular}
\end{table*}

\section*{Appendix 2: Spatial Positions of OJ 287 and the Pulsars, and the Calculations of Parameters $\psi$ and $\mu$}~

The spatial positions of OJ 287 and the 26 pulsars are listed in Table \ref{position}.

\begin{figure}
\begin{center}
\includegraphics[width=0.4\textwidth]{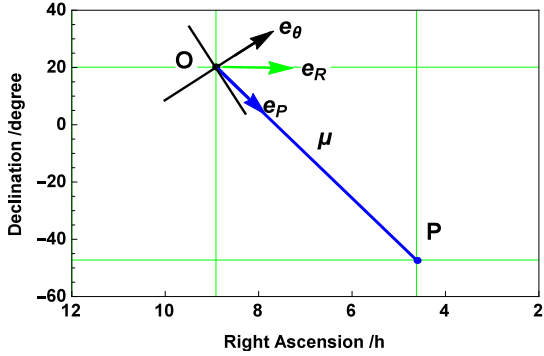}
\caption{The sky locations of OJ287 and PSR J0437-4715.
Point O denotes OJ 287 and P represents PSR J0437-4715.
The black ``cross" on point O is the $(\vec e_{\theta}, \vec e_{\phi})$-basis defined in Section \ref{Strain} , and the black arrow shows the $\vec e_{\theta}$-direction.
The green arrow $\vec e_R$ always points to the prime meridian along the OJ 287-declination line.
The blue curve connecting O and P in fact is not a straight line but a great circular arc, and
the angular distance of arc $\overset{\frown} {OP}$ is $\mu$.
The blue arrow $\vec e_P$ points from O along the arc $\overset{\frown} {OP}$.
$\psi$ is the opening angle between $\vec e_\theta$ and $\vec e_P$,
and $\psi_0$ the angle between $\vec e_\theta$ and $\vec e_R$.
}
\label{sky}
\end{center}
\end{figure}

From the sky locations in the Table \ref{position},  the geometric parameters $\mu$ and $\psi$ of all pulsars can be calculated.
Figure \ref{sky} illustrates the locations of OJ 287 and PSR J0437-4715, as an example, on the celestial sphere.
Note that although Figure \ref{sky} seems on a plane, it is actually on the celestial sphere.

Before calculation, we use the direction angles $(\theta', \phi')$ of spherical coordinate to re-express the sky locations of the pulsars by
\[
\phi'=\frac{\pi}{12} \times {\rm right ascension}, \\
\theta'=\frac{\pi}{2}- \frac{\pi}{180} \times {\rm declination},
\]
in unit of radian.
The prime `` $'$ " here distinguishes  $(\theta', \phi')$ from $(\theta, \phi)$ defined in Section \ref{Strain}.
When the direction angles $(\theta', \phi')$ of OJ 287 and the pulsar are known,
it is not difficult to obtain the angular distance of the great circular arc $\overset{\frown} {OP}$ in Figure \ref{sky},
namely the parameter $\mu$, that
\be
\label{mu}
\mu=\arccos \left[ \cos\theta'_O \cos\theta'_P+\sin\theta'_O \sin\theta'_P\cos(\phi'_P-\phi'_O) \right],
\ee
in unit of radian, where the subscript ``O" is for OJ 287, and ``P" for the pulsar.

Since the direction $\vec e_\theta$ of OJ 287 is totally unknown, the value $\psi$ cannot be undetermined.
However, it is possible to calculate the opening angle between $\vec e_P$ and $\vec e_R$ in Figure \ref{sky},
which is $\psi-\psi_0$ in Table \ref{pulsars}, and take $\psi_0$ as a parameter free from the pulsars.
It is essential to define the signs of the angles $\psi$, $\psi_0$ and $\psi-\psi_0$.
We make an appointment that $\psi$ is the angle by rotating $\vec e_P$ to $\vec e_\theta$, $\psi_0$ by rotating $\vec e_R$ to $\vec e_R$,
$\psi-\psi_0$ by rotating $\vec e_P$ to $\vec e_\theta$ respectively, and when  rotations are anti-clockwise, the corresponding angles are positive.
It is seen that points O, P and the North-Pole point (N) in Figure \ref{sky} form a spherical triangle, with all edges being great circular arcs,
 $\angle ONP=|\phi'_P-\phi'_O|$, and
\[
\angle NOP=\left\{
             \begin{array}{ll}
               \frac{\pi}{2}+\psi-\psi_0, & {\rm case~ 1~ {\rm and}~ case~ 2} \\
               \frac{3\pi}{2}-(\psi-\psi_0), & {\rm case~ 3} ~ , \\
               -(\frac{\pi}{2}+\psi-\psi_0), & {\rm case~ 4}
             \end{array}
           \right.
\]
where case 1 stands for ``$\phi'_P<\phi'_O $ or $\phi'_P>\phi'_O+\pi$ and  $\theta'_P<\theta'_O$",
case 2 for ``$\phi'_P<\phi'_O $ or $\phi'_P>\phi'_O+\pi$ and  $\theta'_P>\theta'_O$",
case 3 for ``$\phi'_O <\phi'_P<\phi'_O+\pi$ and $\theta'_P>\theta'_O$", and
case 4 for ``$\phi'_O <\phi'_P<\phi'_O+\pi$ and $\theta'_P<\theta'_O$".
From the spherical sine theorem, i.e. $\sin(\angle ONP) \cdot \sin(\overset{\frown}{NP})=\sin(\angle NOP) \cdot \sin(\overset{\frown}{OP})$, one obtains
\be
\label{psi}
\psi-\psi_0=\left\{
             \begin{array}{ll}
               -\arccos\left( \frac{\sin|\phi'_O-\phi'_P|\sin\theta'_P}{\sin \mu} \right), & {\rm case~ 1}\\
               \arccos\left( \frac{\sin|\phi'_O-\phi'_P|\sin\theta'_P}{\sin \mu} \right), & {\rm case~ 2}\\
               \pi-\arccos\left( \frac{\sin|\phi'_O-\phi'_P|\sin\theta'_P}{\sin \mu} \right), & {\rm case~ 3} ~ , \\
               \arccos\left( \frac{\sin|\phi'_O-\phi'_P|\sin\theta'_P}{\sin \mu} \right)-\pi, & {\rm case~ 4}
             \end{array}
           \right.
\ee
with unit radian. Taking the value of $\mu$ from Eq. \eqref{mu} into \eqref{psi}, the angle $\psi-\psi_0$ is obtained.

\section*{Appendix 3: Resulting Figures}~

\begin{figure*}
\includegraphics[width=0.9\textwidth]{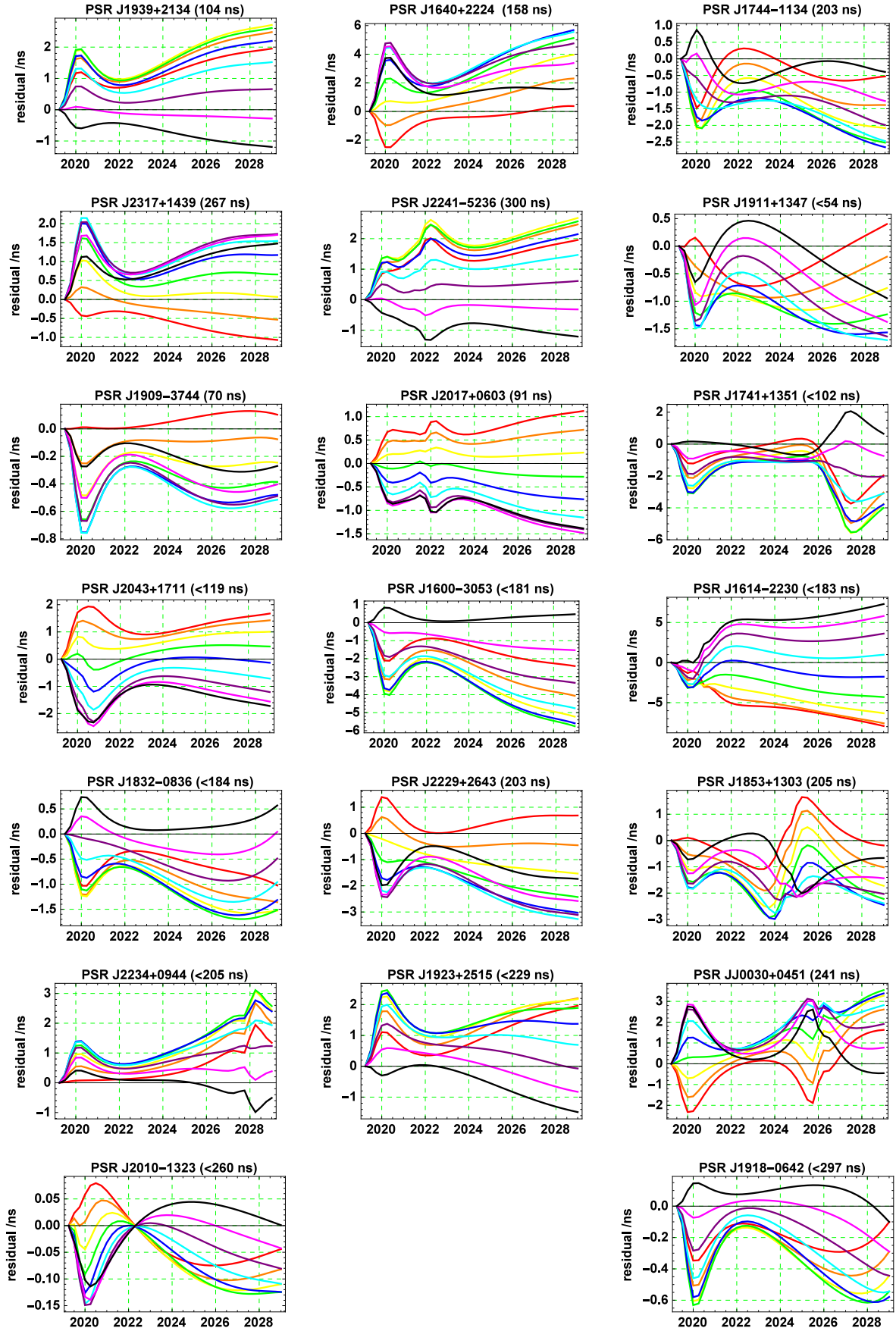}
\caption{The timing residuals for pulsars other from the six ones in Figure \ref{residual}, in the next 10 years.
}
\label{residual2}
\end{figure*}

\begin{figure*}
\begin{center}
\includegraphics[width=0.9\textwidth]{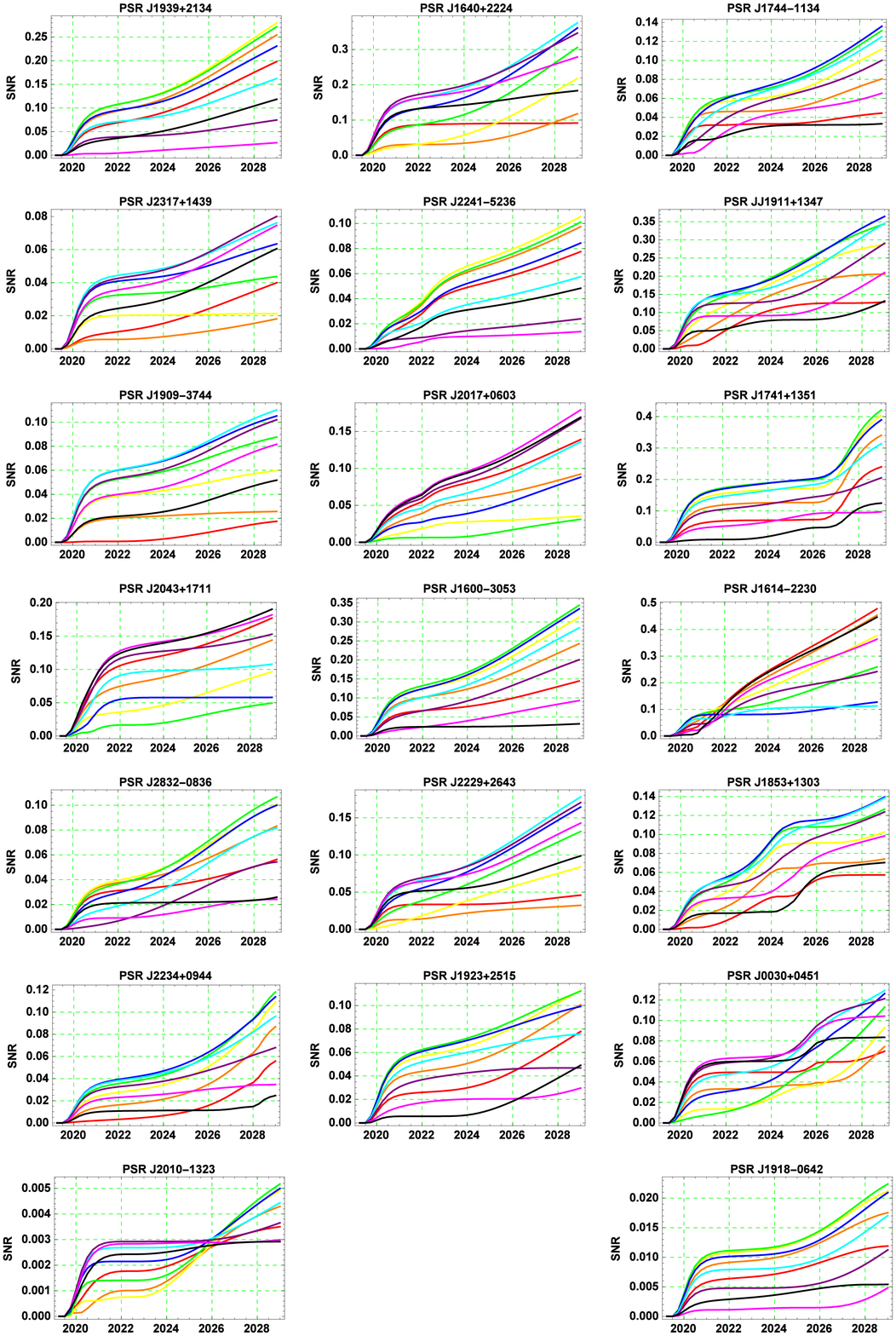}
\caption{The accumulations of individual SNRs for the pulsars other from those in Figure \ref{SNR fig}, in the next 10 years.
}
\label{SNR fig2}
\end{center}
\end{figure*}

This part presents resulting figures which are not given in the text.
Figure \ref{residual2} shows the timing residuals of pulsars , as a complementary to Figure \ref{residual},
and Figure \ref{SNR fig2} shows individual SNRs of pulsars, as a complementary to Figure \ref{SNR fig}.

\label{lastpage}

\bsp
\end{document}